\documentclass{article}
\usepackage[utf8]{inputenc}
\usepackage{authblk}
\usepackage{setspace}
\usepackage[margin=1.25in]{geometry}
\usepackage{graphicx}
\graphicspath{ {./} }
\usepackage{subcaption}
\usepackage{amsmath}

\usepackage{multirow}

\usepackage[style=ieee, 
citestyle=numeric-comp,
sorting=none]{biblatex}
\title{Computing molecular excited states on a D-Wave quantum annealer}

\author[1]{Alexander Teplukhin}
\author[1]{Brian K. Kendrick}
\author[2]{Susan M. Mniszewski}
\author[1]{Yu Zhang}
\author[1]{Ashutosh Kumar}
\author[1]{Christian F. A. Negre}
\author[3]{Petr M. Anisimov}
\author[1$\dagger$]{Sergei Tretiak}
\author[4*]{Pavel A. Dub}

\affil[1]{Theoretical Division, Los Alamos National Laboratory, Los Alamos, NM 87545, USA}
\affil[2]{Computer, Computational and Statistical Sciences Division, Los Alamos National Laboratory, Los Alamos, NM 87545, USA}
\affil[3]{Accelerator Operations and Technology Division, Los Alamos National Laboratory, Los Alamos, NM 87545, USA}
\affil[4]{Chemistry Division, Los Alamos National Laboratory, Los Alamos, NM 87545, USA}
\affil[$\dagger$]{Email: serg@lanl.gov}
\affil[*]{Corresponding author. Email: pdub@lanl.gov}

\date{}

\begin{document}

\maketitle

\begin{abstract}

The possibility of using quantum computers for electronic structure calculations has opened up a promising avenue for computational chemistry. Towards this direction, numerous algorithmic advances have been made in the last five years. The potential of quantum annealers, which are the prototypes of adiabatic quantum computers, is yet to be fully explored. In this work, we demonstrate the use of a D-Wave quantum annealer for the calculation of excited electronic states of molecular systems. These simulations play an important role in a number of areas, such as photovoltaics, semiconductor technology and nanoscience. The excited states are treated using two methods, time-dependent Hartree-Fock (TDHF) and time-dependent density-functional theory (TDDFT), both within a commonly used Tamm-Dancoff approximation (TDA). The resulting TDA eigenvalue equations are solved on a D-Wave quantum annealer using the Quantum Annealer Eigensolver (QAE), developed previously. The method is shown to reproduce a typical basis set convergence on the example H$_2$ molecule and is also applied to several other molecular species. Characteristic properties such as transition dipole moments and oscillator strengths are computed as well. Three potential energy profiles for excited states are computed for NH$_3$ as a function of the molecular geometry. Similar to previous studies, the accuracy of the method is dependent on the accuracy of the intermediate meta-heuristic software called qbsolv.

\end{abstract}


\section{Introduction}

The calculation of electronic states of a molecule has been a routine procedure for theoretical and computational chemistry for decades. While the ground electronic state on its own is undoubtedly the foundation of many chemical theories and methods to describe a large number of physical phenomena, there is a significant amount of chemistry and physics problems where inclusion of excited electronic states is mandatory. These span technological applications, for instance, photovoltaics, lighting and hotocatalysis, and such materials as semiconductors, molecular chromophores, carbon nanotubes, to name a few. A simple example of a phenomenon that requires an excited states description is the absorption of sunlight by pigments in photosynthesis. As a result, by absorbing a photon, the pigment molecule undergoes a transition from the ground to one of the excited states. Understanding these kinds of processes has given rise to a number of modern technologies such photovoltaics, where sunlight is converted into electric current through specifically-designed semiconducting materials \cite{pv1,pv2}. Another related example of an excited-state process is the response of a molecule to an external electromagnetic field. Here, several excited electronic states might become involved in a complex time-dependent dynamics.

One common way to evaluate excited states in electronic structure calculations is time-dependent formulations, such as time-dependent Hartree-Fock (TDHF) and time-dependent density-functional theory (TDDFT) methodologies. TDHF and TDDFT build the excited state description on the ground state self-consistent field (SCF) method and differ in representation of the underlying Hamiltonian and the type of orbitals used (Hartree-Fock or Kohn-Sham). Frequency domain solution of the time-dependent Schr\"{o}dinger equation in either of these formulations leads to the random-phase approximation (RPA) non-Hermitian eigenvalue problem \cite{rpa1,rpa2,rpa3,rpa4}, which can be further simplified to a Hermitian eigenvalue problem by means of the commonly-used Tamm-Dancoff approximation (TDA) in TDDFT \cite{tda1,tda2,tda3}. This is formally equivalent to the configuration interaction with single substitutions (CIS), a term frequently used in the wave function theories. At this point, the only pragmatic concern is to find an efficient eigensolver for the RPA or CIS matrix.

In parallel to the advances in electronic structure theory, quantum computers, that promise exponentially faster calculations, become more mature every year. The question that has emerged is how to map existing chemistry problems, such as excited state calculations, to this new type of computing hardware. Whereas a solution has already been found for the gate-based quantum computers in the form of the Variational Quantum Eigensolver (VQE) \cite{vqe1,vqe2,vqe3}, there has not been much progress in the direction of quantum annealers, D-Wave quantum annealers specifically. In the past, two methods \cite{purdue,lumionics} were proposed to solve general configuration interaction (CI) problems on quantum annealers, both starting from the second-quantized formulation of the electronic problem.

Previously, we developed the Quantum Annealer Eigensolver (QAE), which is capable of computing a few eigenvalues and eigenvectors of a given real symmetric or complex Hermitian matrix on a D-Wave quantum annealer (non-Hermitian matrices are supported in a limited way). The QAE was successfully applied to vibrational, scattering and electronic structure problems \cite{qae1,qae2,qae3,qae4}, and was recently improved and applied to lattice gauge theory \cite{aqae}. While the last study \cite{qae4} on electronic structure covered a significant number of molecules, all matrices were general CI matrices.

In this work, we extend our annealer-based electronic structure calculations to the time-dependent TDDFT and TDHF formulations and target excited states specifically. In the Results section, we briefly review the methodology and setup, and demonstrate the convergence of excitation energy with basis set for the hydrogen molecule. Calculations of excitation energies for a number of small molecules are then presented followed by evaluation of other properties of interest such as transition dipole moments and oscillator strengths. We conclude the Results section with the excitation energies of NH$_3$ molecule reported as a function of the nuclear geometry (umbrella inversion). The Discussion section elaborates on the accuracy of the results obtained and outlines the limitations of the proposed method. The electronic structure code, the QAE algorithm and D-Wave quantum annealer are described in the Materials and Methods section.

\section{Results}

\subsection{Method overview and setup}
 
The method is based on using a standard electronic structure code, such as PySCF \cite{pyscf1,pyscf2}, where the eigensolver is replaced with the QAE \cite{qae1,qae2,qae3}. The QAE generates quadratic unconstrained binary optimization (QUBO) problems, solvable by a D-Wave quantum annealer. To solve a large QUBO problem, the qbsolv software \cite{qbsolv} is used to iteratively minimize the QUBO function by decomposing it into smaller pieces, called subQUBOs, that fit on a quantum annealer. Since the qbsolv is a heuristic QUBO solver, the resulting solution to the QUBO problem is non-exact, which makes the whole procedure approximate as a consequence. All eigenvalue problems in this study are real-valued TDA eigenvalue problems, either TDHF or TDDFT type. The chosen exchange–correlation energy functional in the TDDFT calculations is B3LYP \cite{b3lyp1,b3lyp2,b3lyp3}. In what follows, we present the results for the TDDFT case only. The TDHF results are reported in the Supplementary Materials.

\subsection{Excitation energy convergence with basis set}

\begin{figure}
    \centering
    \includegraphics[scale=0.6]{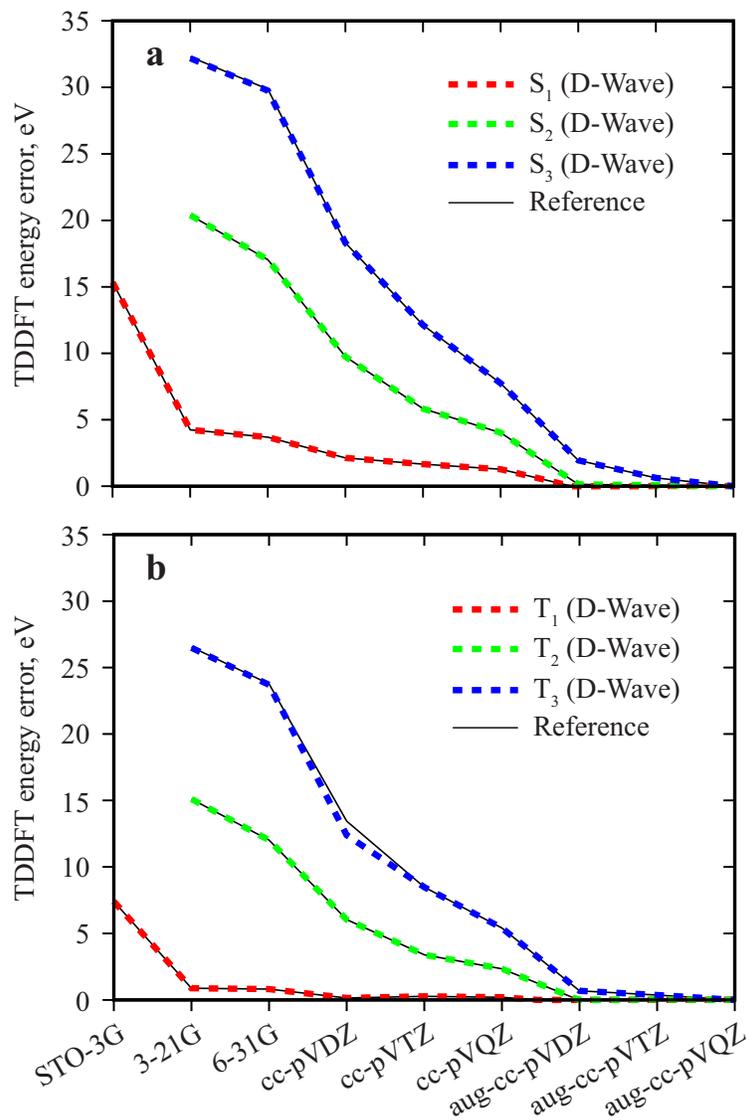}
    \caption{Convergence of TDDFT excitation energies for the H$_2$ molecule with respect to basis set. Singlet (a) and triplet (b) calculations are shown separately. The first three excitation energies are computed using PySCF with the QAE and D-Wave 2000Q (dashed red, green and blue curves) and using the unmodified PySCF (thin black curves). The energy error is given relative to the aug-cc-pVQZ reference calculation.}
    \label{fig:1}
\end{figure}

Figure~\ref{fig:1} shows the convergence of the excitation energy with basis set. In this study, the first three excitation energies of the hydrogen molecule were computed using 9 basis sets, from the smallest STO-3G with the TDA matrix size of 1x1 (for which the solution is trivial) to the largest aug-cc-pVQZ with the TDA matrix size of 91x91. Similar to previous work \cite{qae2}, we use $K = 10$ binary variables to represent one eigenvector element, which translates to a QUBO problem of size that ranges from 10 to 910 variables. The respective excitations to the singlet states (S$_1$, S$_2$ and S$_3$) and the triplet states (T$_1$, T$_2$ and T$_3$) are treated as two independent calculations. As can be seen in Fig.~\ref{fig:1}, the PySCF calculation that utilizes the QAE and D-Wave 2000Q quantum annealer closely agrees with the result of the reference calculation, which was done using the unmodified PySCF code on a classical computer. The unmodified PySCF relies on the efficient Davidson algorithm to compute eigenvalues and eigenvectors.

The largest deviation of 1 eV is observed for the T$_3$ excited state in the cc-pVDZ basis set. Such fluctuations are fortuitous and typically are caused by the heuristic nature of qbsolv \cite{qae2}. Specifically for the QAE, which computes the eigenpairs sequentially, the qbsolv error propagates and accumulates from small eigenvalues to large eigenvalues. As a result, the third excitation energy tends to be less accurate than the first and second excitation energies. All plotted energy errors are given relative to the rightmost reference calculation in the figure, i.e, in aug-cc-pVQZ basis. The internuclear distance between the two hydrogens were taken from the NIST database \cite{nist}, individually for each basis set.

\subsection{Excitation energy of different molecular species}

The excitation energies of different molecular species are given in Tables~\ref{tab:1} and \ref{tab:2} for singlet and triplet states, respectively. The calculations include species with up to eight atoms, however, it should be possible to target even larger systems with the present method. The electronic structure theory is TDDFT where B3LYP functional is coupled with 6-31G basis set. The equilibrium geometries were taken from the NIST database \cite{nist}. The resultant TDA matrix sizes are given in the second column of the table and we use $K=10$ qubits to discretize eigenvector coefficients.

\begin{table}
    \caption{Singlet TDDFT excitation energies (eV).}
    \begin{tabular}{ccccccccccc}
        \hline
        \multirow{2}{*}{Molecule} & \multirow{2}{*}{Mat. size} & \multicolumn{3}{c}{Reference$^*$} & \multicolumn{3}{c}{QAE (D-Wave)$^\dagger$} & \multicolumn{3}{c}{Error$^\ddag$} \\ \cline{3-5} \cline{6-8} \cline{9-11}
        & & S$_1$ & S$_2$ & S$_3$ & S$_1$ & S$_2$ & S$_3$ & S$_1$ & S$_2$ & S$_3$ \\
        \hline
        H$_2$        &   3 & 15.157 & 28.645 & 43.090 & 15.157 & 28.645 & 42.975 & 0.000 & 0.000 & -0.115 \\
        H$_3^+$      &   5 & 20.446 & 20.446 & 32.241 & 20.446 & 20.446 & 32.241 & 0.000 & 0.000 &  0.000 \\
        HF           &  30 &  9.857 &  9.857 & 15.405 &  9.860 &  9.862 & 15.420 & 0.003 & 0.005 &  0.015 \\
        BeH$_2$      &  30 &  6.420 &  6.420 &  8.422 &  6.420 &  6.421 &  8.422 & 0.000 & 0.001 &  0.000 \\
        H$_2$O       &  40 &  7.896 &  9.649 & 10.044 &  7.900 &  9.664 & 10.051 & 0.004 & 0.015 &  0.007 \\
        NH$_3$       &  50 &  7.140 &  9.186 &  9.186 &  7.147 &  9.194 &  9.196 & 0.006 & 0.008 &  0.010 \\
        H$_2$S       &  72 &  6.617 &  7.221 &  9.939 &  6.618 &  7.235 &  9.940 & 0.001 & 0.014 &  0.001 \\
        HOCl         & 143 &  3.472 &  4.708 &  6.321 &  3.482 &  4.731 &  6.345 & 0.010 & 0.024 &  0.024 \\
        C$_2$H$_6$   & 189 & 10.989 & 10.989 & 11.571 & 10.996 & 10.998 & 11.601 & 0.006 & 0.009 &  0.030 \\
        CH$_2$Cl$_2$ & 378 &  6.483 &  6.706 &  6.854 &  6.496 &  6.723 &  6.864 & 0.014 & 0.017 &  0.009 \\
        \hline
        \end{tabular}
    \label{tab:1} \\
    $^*$       Reference calculation on CPU using unmodified PySCF \\
    $^\dagger$ PySCF was modified to use the QAE and D-Wave 2000Q \\
    $^\ddag$   Difference between the two types of calculation
\end{table}

\begin{table}
    \caption{Triplet TDDFT excitation energies (eV).}
    \begin{tabular}{ccccccccccc}
        \hline
        \multirow{2}{*}{Molecule} & \multirow{2}{*}{Mat. size} & \multicolumn{3}{c}{Reference$^*$} & \multicolumn{3}{c}{QAE (D-Wave)$^\dagger$} & \multicolumn{3}{c}{Error$^\ddag$} \\ \cline{3-5} \cline{6-8} \cline{9-11}
        & & T$_1$ & T$_2$ & T$_3$ & T$_1$ & T$_2$ & T$_3$ & T$_1$ & T$_2$ & T$_3$ \\
        \hline
        H$_2$        &   3 & 10.978 & 23.129 & 36.199 & 10.978 & 23.129 & 36.131 & 0.000 & 0.000 & -0.068 \\
        H$_3^+$      &   5 & 15.532 & 15.532 & 28.900 & 15.532 & 15.532 & 28.900 & 0.000 & 0.000 &  0.000 \\
        HF           &  30 &  9.104 &  9.104 & 12.250 &  9.106 &  9.110 & 12.255 & 0.001 & 0.005 &  0.006 \\
        BeH$_2$      &  30 &  5.872 &  5.872 &  6.566 &  5.872 &  5.873 &  6.566 & 0.000 & 0.001 &  0.000 \\
        H$_2$O       &  40 &  7.101 &  8.555 &  9.510 &  7.101 &  8.565 &  9.510 & 0.000 & 0.010 &  0.000 \\
        NH$_3$       &  50 &  6.278 &  8.600 &  8.600 &  6.280 &  8.601 &  8.602 & 0.002 & 0.001 &  0.003 \\
        H$_2$S       &  72 &  6.058 &  6.312 &  8.047 &  6.058 &  6.312 &  8.078 & 0.000 & 0.000 &  0.030 \\
        HOCl         & 143 &  2.356 &  3.516 &  5.358 &  2.372 &  3.546 &  5.469 & 0.016 & 0.030 &  0.111 \\
        C$_2$H$_6$   & 189 & 10.350 & 10.350 & 10.844 & 10.388 & 10.491 & 10.911 & 0.038 & 0.141 &  0.068 \\
        CH$_2$Cl$_2$ & 378 &  5.840 &  5.853 &  6.174 &  5.860 &  6.098 &  6.329 & 0.020 & 0.245 &  0.155 \\
        \hline
        \end{tabular}
    \label{tab:2} \\
    $^*$       Reference calculation on CPU using unmodified PySCF \\
    $^\dagger$ PySCF was modified to use the QAE and D-Wave 2000Q \\
    $^\ddag$   Difference between the two types of calculation
\end{table}

Similar to the basis set convergence study, we compare the calculations using the QAE and the D-Wave 2000Q to a reference calculation using the unmodified PySCF on a classical computer. The errors given in the last three columns of the tables tend to be very small for small molecules and tend to increase for larger molecules. This is expected because, in general, heuristic approaches work quite well for small problems and give an approximate result for large problems. Nevertheless, all errors are well below an accuracy of 0.2 to 0.3 eV, tentatively characterizing performance of B3LYP model for real life molecular systems. The largest deviation (0.24 eV) is observed for the T$_2$ energy for CH$_2$Cl$_2$, which we will examine in more detail in the Discussion section. Due to the heuristic nature of qbsolv, the errors are not deterministic and fluctuate from run to run.

One might notice that the third excitation energy of the hydrogen molecule for both singlets and triplets is non-monotonic with the excitation number. For instance, the error for the S$_3$ state is large and negative while the matrix size is small, 3x3. From our experience, this seems to be a special case for the QAE. We find that the largest eigenvalue, computed using the QAE, always has an additional error, besides the accumulated error from smaller eigenvalues. This might be related to the fact that, at this point in the calculation, all eigenpairs except the last one were shifted outside of the dynamic range of the initial matrix and the last eigenpair sits ``isolated''. Since we target three eigenvalues in the present study, the hydrogen case with its 3x3 matrix is the only case where the last (largest) eigenpair of the matrix needs to be calculated. In all other cases, the matrices are larger than 3x3 and the last eigenvalue is never computed. All errors are reasonable for those matrices.

Since QAE computes both eigenvalues (excitation energies) and eigenvectors (wave function expansion coefficients), it becomes possible to compute different properties of electronic transitions. For example, Table~\ref{tab:3} shows the computed transition dipole moments (TDMs) in atomic units for singlet states. These dipole moments are then used to compute the respective oscillator strengths, Table~\ref{tab:4}, which characterize the ability of absorption or emission of electromagnetic radiation in transitions between energy levels of a molecule. As can be seen, the PySCF code aided with the QAE and D-Wave 2000Q is capable of reproducing these properties quite accurately, at least for the molecular species studied. The triplet state TDMs and oscillator strength have not been computed, simply because those are strictly vanishing in the absence of spin-orbit coupling. Alpha and beta spin contributions cancel each other out for triplet states.

\begin{table}
    \caption{Singlet TDDFT transition dipole moments (au).}
    \begin{tabular}{ccccccccccc}
        \hline
        \multirow{2}{*}{Molecule} & \multirow{2}{*}{Mat. size} & \multicolumn{3}{c}{Reference$^*$} & \multicolumn{3}{c}{QAE (D-Wave)$^\dagger$} & \multicolumn{3}{c}{Error$^\ddag$} \\ \cline{3-5} \cline{6-8} \cline{9-11}
        & & S$_1$ & S$_2$ & S$_3$ & S$_1$ & S$_2$ & S$_3$ & S$_1$ & S$_2$ & S$_3$ \\
        \hline
        H$_2$        &   3 & 1.974 & 0.000 & 0.133 & 1.974 & 0.000 & 0.207 &  0.000 &  0.000 &  0.073 \\
        H$_3^+$      &   5 & 1.453 & 1.453 & 0.000 & 1.453 & 1.453 & 0.000 &  0.000 &  0.000 &  0.000 \\
        HF           &  30 & 0.038 & 0.038 & 1.030 & 0.038 & 0.037 & 1.027 &  0.000 & -0.001 & -0.004 \\
        BeH$_2$      &  30 & 0.000 & 0.000 & 2.447 & 0.000 & 0.000 & 2.447 &  0.000 &  0.000 &  0.000 \\
        H$_2$O       &  40 & 0.062 & 0.374 & 0.000 & 0.063 & 0.352 & 0.000 &  0.001 & -0.023 &  0.000 \\
        NH$_3$       &  50 & 0.058 & 0.051 & 0.051 & 0.057 & 0.054 & 0.061 &  0.000 &  0.003 &  0.010 \\
        H$_2$S       &  72 & 0.000 & 0.016 & 0.752 & 0.000 & 0.015 & 0.749 &  0.000 & -0.001 & -0.002 \\
        HOCl         & 143 & 0.004 & 0.037 & 0.024 & 0.003 & 0.029 & 0.031 & -0.001 & -0.009 &  0.006 \\
        C$_2$H$_6$   & 189 & 0.000 & 0.000 & 0.000 & 0.000 & 0.000 & 0.000 &  0.000 &  0.000 &  0.000 \\
        CH$_2$Cl$_2$ & 378 & 0.009 & 0.221 & 0.000 & 0.011 & 0.236 & 0.000 &  0.002 &  0.015 &  0.000 \\
        \hline
        \end{tabular}
    \label{tab:3} \\
    $^*$       Reference calculation on CPU using unmodified PySCF \\
    $^\dagger$ PySCF was modified to use the QAE and D-Wave 2000Q \\
    $^\ddag$   Difference between the two types of calculation
\end{table}

\begin{table}
    \caption{Singlet TDDFT oscillator strengths (unitless).}
    \begin{tabular}{ccccccccccc}
        \hline
        \multirow{2}{*}{Molecule} & \multirow{2}{*}{Mat. size} & \multicolumn{3}{c}{Reference$^*$} & \multicolumn{3}{c}{QAE (D-Wave)$^\dagger$} & \multicolumn{3}{c}{Error$^\ddag$} \\ \cline{3-5} \cline{6-8} \cline{9-11}
        & & S$_1$ & S$_2$ & S$_3$ & S$_1$ & S$_2$ & S$_3$ & S$_1$ & S$_2$ & S$_3$ \\
        \hline
        H$_2$        &   3 & 0.733 & 0.000 & 0.141 & 0.733 & 0.000 & 0.218 & 0.000 &  0.000 &  0.077 \\
        H$_3^+$      &   5 & 0.728 & 0.728 & 0.000 & 0.728 & 0.728 & 0.000 & 0.000 &  0.000 &  0.000 \\
        HF           &  30 & 0.009 & 0.009 & 0.389 & 0.009 & 0.009 & 0.388 & 0.000 &  0.000 & -0.001 \\
        BeH$_2$      &  30 & 0.000 & 0.000 & 0.505 & 0.000 & 0.000 & 0.505 & 0.000 &  0.000 &  0.000 \\
        H$_2$O       &  40 & 0.012 & 0.089 & 0.000 & 0.012 & 0.083 & 0.000 & 0.000 & -0.005 &  0.000 \\
        NH$_3$       &  50 & 0.010 & 0.011 & 0.011 & 0.010 & 0.012 & 0.014 & 0.000 &  0.001 &  0.002 \\
        H$_2$S       &  72 & 0.000 & 0.003 & 0.183 & 0.000 & 0.003 & 0.182 & 0.000 &  0.000 & -0.001 \\
        HOCl         & 143 & 0.000 & 0.004 & 0.004 & 0.000 & 0.003 & 0.005 & 0.000 & -0.001 &  0.001 \\
        C$_2$H$_6$   & 189 & 0.000 & 0.000 & 0.000 & 0.000 & 0.000 & 0.000 & 0.000 &  0.000 &  0.000 \\
        CH$_2$Cl$_2$ & 378 & 0.002 & 0.036 & 0.000 & 0.002 & 0.039 & 0.000 & 0.000 &  0.003 &  0.000 \\
        \hline
        \end{tabular}
    \label{tab:4} \\
    $^*$       Reference calculation on CPU using unmodified PySCF \\
    $^\dagger$ PySCF was modified to use the QAE and D-Wave 2000Q \\
    $^\ddag$   Difference between the two types of calculation
\end{table}

\subsection{Excitation energy as a function of molecular geometry}

\begin{figure}
    \centering
    \includegraphics[scale=0.6]{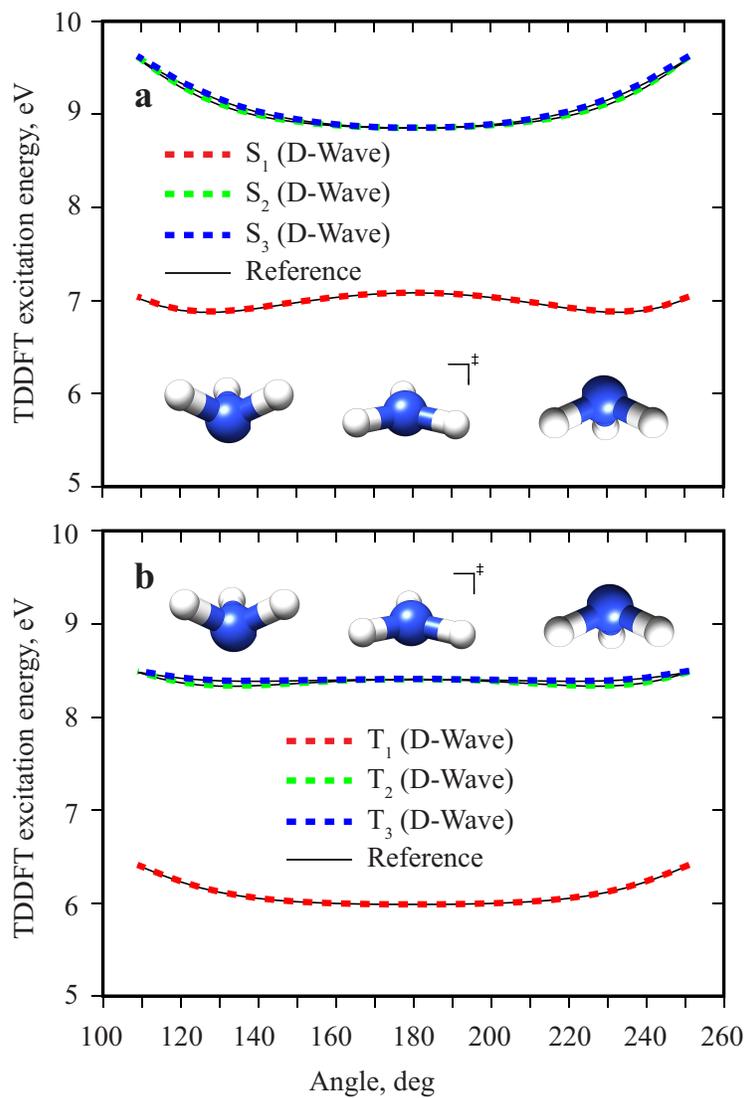}
    \caption{TDDFT excitation energies for the umbrella inversion of ammonia (NH$_3$). Singlet (a) and triplet (b) calculations are shown separately. The first three excitation energies are computed using the QAE and D-Wave 2000Q quantum annealer (dashed red, green and blue curves). The reference calculation is shown for comparison (thin black curves). The angle is the H-N-H-H dihedral angle.}
    \label{fig:2}
\end{figure}

It is often necessary to know how the excited state energies or excitation energies vary with molecular geometry. Such data is the determining component in molecular dynamics studies that involve multiple electronic states. The method presented in this paper can be used out-of-the-box to compute excited state potential energy profiles \cite{exstmd}. Figure~\ref{fig:2} shows the first three excitation energies of ammonia (NH$_3$) as a function of the H-N-H-H dihedral angle. The geometry change represents the famous ``umbrella'' inversion of NH$_3$. The QAE+D-Wave computed excitation energies agree very well with the reference calculation on a classical computer, for both singlets and triplets. Importantly, the QAE was capable of handling the almost degenerate pair of the second and third excited states along the whole angle range shown and all three resultant energy curves are very smooth.

\subsection{CIS calculations}

All the TDDFT-TDA calculations presented in this study were also repeated with the TDHF-TDA (CIS) theory and are given in the Supplementary Material, Part 1. Supplemental Figures S1 and S2 demonstrate the basis set convergence for the hydrogen molecule and excitation energies for ammonia inversion, respectively. The excitation energies, TDMs and oscillator strengths of different molecular species are given in Supplemental Tables S1-S4.

Overall, we find that the method works for the CIS calculations too. However, the QAE (D-Wave) errors tend to be larger than those in the TDDFT calculations, sometimes by one to two orders of magnitude. For example, the S$_3$ and T$_3$ excitation energy errors for the last three molecules in Supplemental Tables~S1 and S2 are about 2 to 3 eV for the CIS case, which is an order of magnitude larger than the QAE errors in TDDFT calculations shown earlier, although the TDA matrix size is the same for both TDDFT and TDHF. Typically, a large error implies that the QAE has converged to the wrong excited state, which can be checked with a wave function overlap analysis, see Table~\ref{tab:5}. Also, the less accurate CIS result implies that the QAE (and potentially qbsolv and an annealer) is sensitive to the problem representation and the TDDFT B3LYP formalism seems to be more favorable for the QAE. The fluctuations in the third excitation energy of ammonia in Figure S2 also indicate that the QAE struggles with closely degenerate states in the CIS case.

\subsection{Purely classical solution of QUBOs}

While the main focus of the paper is to implement the time-dependant SCF methods on a D-Wave quantum annealer, all of the presented calculations were also done in a purely classical mode, for debugging purposes primarily. In this mode, we still use PySCF, QAE and qbsolv, but all subQUBO problems are minimized classically on a CPU using the Tabu search. The latter is an efficient local search technique that discourages the search from coming back to previously-visited solutions \cite{tabu}. The qbsolv uses this method by default. All of the results from the classical calculations are given in Part 2 of the Supplementary Material. There is not much difference between using a quantum annealer or Tabu search for subQUBOs, and we include the classical results for completeness only. Additionally, we find that the completely classical solution of QUBOs is much faster than the use of the D-Wave annealer for this task. In other words, there is no quantum advantage. The TDDFT calculations still appear to be much more accurate than CIS calculations.

Lastly, to demonstrate that all of the results are not unique and may fluctuate from run to run due to the heuristic nature of qbsolv, we performed 10 identical QAE calculations for the C$_2$H$_6$ molecule in the smallest STO-3G basis set (TDA matrix size 63x63). Tables~S13 and S14 in the Part 3 of the Supplementary Materials show that the excitation energies change between runs. Averaging over all 10 runs gives the QAE errors 0.004, 0.130 and 0.173 eV for the first three singlets and 0.022, 0.340 and 1.325 eV for the first three triplets. These are purely classical calculations, where subQUBOs were minimized classically on the CPU.

\section{Discussion}

Although the TDDFT energy errors reported in Tables~\ref{tab:1} and \ref{tab:2} appear to be quite acceptable compared to a typical accuracy of TDDFT methods, there is at least one case which might require an additional investigation. This case is the triplet state calculation for the CH$_2$Cl$_2$ molecule. While the QAE error of the first excitation energy T$_1$ is quite accurate, i.e., 0.02 eV, the errors of the second and third excitation energies are an order of magnitude larger, although still near the acceptable TDDFT accuracy of 0.2 eV.

To investigate this sudden increase in the QAE error, we examined the computed eigenvectors, i.e., wave function expansion coefficients. To quantify the degree of similarity between the eigenvectors computed using the QAE on the D-Wave and eigenvectors from the reference calculation, we calculated the overlaps, see Table~\ref{tab:5}. In the table, the first five excitation energies from the reference calculation and the first three QAE excitation energies are given in the first two rows. The third row shows the overlaps between the QAE and reference eigenvectors for T$_1$, T$_2$ and T$_3$. It should be noted that, technically, the maximum overlap for the TDDFT wave functions is 0.5, because there are two types of electrons in the theory, alpha (spin up) and beta (spin down) electrons. In the present analysis, we multiply all overlaps by the factor of 2 for convenience, so that the maximum (ideal) overlap is one. One can see that the QAE T$_1$ state has converged correctly, since the overlap 0.997 is very close to one. In contrast, the QAE T$_2$ state shows much less similarity with the reference T$_2$ eigenvector, whereas QAE T$_3$ is completely incorrect and has zero overlap with the reference T$_3$. In order to determine what states the QAE has converged to, we calculated the overlaps between the QAE T$_2$ state and all five states from the reference calculation and then did the same for T$_3$, see the fourth and fifth rows in the table. We find that the QAE T$_2$ state bears some similarity with both T$_2$ and T$_3$ from the reference calculation, the overlap values being 0.641 and 0.370, respectively, whereas the QAE T$_3$ state ended up being the T$_4$ state, with the overlap of 0.999. The QAE excitation energies also confirm this observation: the QAE T$_2$ energy of 6.098 eV is between the T$_2$ and T$_3$ reference energies, 5.853 eV and 6.174 eV, respectively, whereas the QAE T$_3$ energy of 6.329 eV is very close to the T$_4$ reference energy of 6.312 eV. The overlap analysis shows that the QAE + qbsolv method may not be able to resolve some of the states, especially when closely degenerate states are present in a particular molecular system. This is not too surprising, because degeneracy may occasionally raise a problem for classical iterative eigensolvers as well.

\begin{table}
    \centering
    \caption{Overlap analysis for triplet states of CH$_2$Cl$_2$.}
    \begin{tabular}{cccccc}
        \hline  
        & T$_1$ & T$_2$ & T$_3$ & T$_4$ & T$_5$ \\
        \hline
        Ref. energy (eV)      & 5.840 & 5.853 & 6.174 & 6.312 & 7.280 \\
        QAE energy (eV)       & 5.860 & 6.098 & 6.329 &     - &     - \\
        Overlaps$^*$           & 0.997 & 0.641 & 0.000 &     - &     - \\
        QAE T$_2$ overlap$^*$ & 0.043 & 0.641 & 0.370 & 0.001 & 0.001 \\
        QAE T$_3$ overlap$^*$ & 0.000 & 0.001 & 0.000 & 0.999 & 0.015 \\
        \hline
        \end{tabular}
    \label{tab:5} \\
    $^*$ See text for the description of overlaps \\
\end{table}

There are several ways to improve the method and many of them have been discussed previously \cite{qae1,qae2,qae3}. For example, by replacing the open-source qbsolv \cite{qbsolv} with the Qatalyst QUBO solver (previously known as Mukai QUBO solver) one can reduce the QAE errors by one to two orders of magnitude \cite{qae4}. In the present work, we have already improved the choice of the shift in the QAE that is used to compute several eigenvalues, see Methods and Materials section.

Another important aspect is the initial guess for the QAE. Choosing a better guess might help decrease the QAE errors and reduce the chance for missing states in the case of degeneracies. For classical iterative eigensolvers, it was shown that the quality of the initial guess for the excitation vector transition density matrix is critical for improved convergence and stability \cite{tretiak}. Unfortunately, it is not possible to specify the initial guess for the QAE right now, simply because the underlying qbsolv does not support user-specified inputs and generates guesses randomly. Besides the initial guess, the method can also be enhanced by using of subspace iterative techniques, such as the Davidson algorithm. With the subspace construction, the matrices that need to be solved for eigenvalues are expected to be smaller, which may improve the accuracy of the whole procedure as a consequence.

\section{Conclusion}

In this work, the time-dependent electronic structure methods  for calculating electronically excited states have been implemented on modern D-Wave quantum annealers by replacing the eigensolver. The approach relies on using the Quantum Annealer Eigensolver (QAE) to convert the time-dependent Hartree-Fock (TDHF) and time-dependent density-functional theory (TDDFT) eigenvalue problems to quadratic unconstrained binary optimization (QUBO) problems, solvable by D-Wave quantum annealers. The auxiliary software qbsolv \cite{qbsolv} is used to solve large QUBO problems, that do not fit on D-Wave quantum annealers. We show that the proposed method is capable of reproducing the results of classical (reference) calculations, which include the basis set convergence of excitation energies of the hydrogen molecule, the excitation energies of different molecular species up to eight atoms and excited state potential energy surfaces for the example of umbrella inversion of ammonia. All QAE excitation energy errors are less than or about 0.2 eV for the TDDFT calculations, which is an acceptable accuracy of the TDDFT method in general. In contrast, we find that the TDHF(CIS) + QAE errors tend to be one to two orders of magnitude larger than TDDFT+QAE errors. In future work, a more robust treatment of degenerate states is needed which should include an option to specify the initial guess for both the QAE and the qbsolv. Thus, we hope the results of this work will encourage researchers to develop qbsolv or similar solvers with the possibility to specify the initial guess in lieu of a random one.

As mentioned earlier, the current approach is not exact, primarily due to the fact that the qbsolv is not an exact QUBO solver. However, more importantly, the methodology demonstrates how a time-dependent electronic structure formalism can be mapped to modern D-Wave quantum annealers. When more advanced quantum annealers become available, i.e. the ones supporting operators beyond $\sigma_z$ and having more advanced connectivity, alternative and potentially more accurate techniques could be utilized to estimate the energy gaps of the target (electronic) Hamiltonians \cite{matsuzaki}. Currently, there is no quantum advantage, because the calculations involving the QPU are much slower than both the classical solution of QUBO problems on the CPU and the classical diagonalization on the CPU. However, one might have an advantage regarding the scaling in the future when more advanced quantum annealers become available.

\section{Materials and Methods}

\subsection{Electronic structure code}

The electronic structure code of choice is PySCF \cite{pyscf1,pyscf2}, because it is written in Python (same as the D-Wave client software) and is relatively easy to modify. As such, we added a switch to choose either the Davidson eigensolver, used by default in PySCF, or the QAE. Since the QAE is a matrix-based algorithm, i.e. it takes a matrix on input, a bootstrapping code was added to the electronic structure code where a TDA matrix is explicitly constructed using the PySCF's matrix-vector function and the standard basis (also called natural basis), where all vector coordinates are zero, except one that equals 1. Both eigensolvers compute a few smallest eigenvalues, or excitation energies, and associated eigenvectors, which are later used in the calculation of transition dipole moments and oscillator strengths. All reference calculations are performed with an unmodified PySCF with the default Davidson eigensolver. The details about TDA implementations can be found elsewhere \cite{tda1,tda2,tda3}. The efficient classical algorithms for both RPA and TDA are analyzed and benchmarked in the relevant literature \cite{tretiak}.

\subsection{Quantum Annealer Eigensolver}

A few lowest eigenvalues and eigenvectors of a TDA matrix are computed using the Quantum Annealer Eigensolver (QAE) \cite{qae1,qae2,qae3}. The QAE represents an eigenvalue problem with a number of quadratic unconstrained binary optimization (QUBO) problems, which are solved on a D-Wave quantum annealer. The QUBO problem is a problem of minimizing a quadratic polynomial $E(\mathbf{x})$ over binary variables $x_i$. The polynomial is given by
\begin{eqnarray}
E(\mathbf{x})=\sum_{i=1}^N \sum_{j=1}^i x_i Q_{ij} x_j, \label{eq:qubo}
\end{eqnarray}
where $\mathbf{Q}$ is a matrix of coefficients and the vector $\mathbf{x}$ is a binary string. The $\mathbf{x}_{opt}$ which minimizes $E(\mathbf{x})$ is the solution to the QUBO.

To find the smallest eigenvalue, the QAE searches for the minimum of the Rayleigh-Ritz quotient (RRQ) 
\begin{eqnarray}
R_A(\mathbf{v})=\mathbf{v}^*\mathbf{A}\mathbf{v}/\mathbf{v}^*\mathbf{v} \label{eq:rrq}
\end{eqnarray}
where $\mathbf{A}$ is a given matrix and $\mathbf{v}$ is an arbitrary vector. For the normalized $\mathbf{v}$, the quotient is $R_A(\mathbf{v})=\mathbf{v}^*\mathbf{A}\mathbf{v}$ and the matrix $\mathbf{A}$ can be mapped to the QUBO matrix $\mathbf{Q}$ by representing elements of $\mathbf{v}$ using binary variables $\mathbf{x}$. An additional term responsible for the vector normalization $\mathbf{v}^*\mathbf{v}$ is added to the QUBO, giving the final objective function
\begin{eqnarray}
F(\mathbf{v})=\mathbf{v}^*\mathbf{A}\mathbf{v}+\lambda\mathbf{v}^*\mathbf{v} . \label{eq:f}
\end{eqnarray}
The strength of the second term, i.e., vector normalization penalty $\lambda$, balances two components, the energy and the norm, and is not known a priori. In the QAE, the penalty $\lambda$ is found iteratively which necessitates solution of multiple QUBO problems.

To obtain several eigenpairs, the previously calculated ones are simply shifted higher in the eigenspectrum and the procedure is repeated. A few of the smallest eigenvalues of the TDA matrix are the sought-for excitation energies. Since the resulting QUBO problems are an order of magnitude larger than the fully-connected graph of qubits on a D-Wave quantum annealer, qbsolv \cite{qbsolv} is utilized to decompose the problem into smaller subQUBOs. The convergence of the eigenvalues with respect to the number of qubits used to represent an eigenvector element was thoroughly studied in the previous paper for different matrix sizes \cite{qae2}. Thus, the number of qubits used in the present work is the same, $K=10$. Since an eigenvector element is represented as a linear combination of $K$ qubits, the expectation values of quantum-mechanical operators, such as the Hamiltonian or dipole moment operator, converge quadratically with $K$. The reader is encouraged to familiarize themselves with more details about the QAE algorithm in the Methods section of a recent QAE paper \cite{qae2}. 

The only change to the QAE algorithm in the present work is the implementation of the Gershgorin circle theorem to obtain the upper bound on the difference between the largest and the smallest eigenvalues of a matrix. The dynamic range estimate is then used to shift previously computed eigenpairs higher in the eigenspectrum in order to compute the next eigenpairs. We believe that the new shift is more universal and accurate than the manually-scaled difference between the maximum and minimum elements of a given matrix, used previously.

\subsection{D-Wave quantum annealer}

LANL’s D-Wave 2000Q was used in all hardware-mode calculations. The reader is referred to the Methods section of a recent QAE paper \cite{qae2} for details about using this particular system. The only procedural change in this work is that the minor-embedding is now computed in advance and is reused for the minimization of all subQUBOs. This change reduced the total computational time significantly.

\subsection*{Conflicts of Interest}
The authors declare that there is no conflict of interest regarding the publication of this article.

\subsection*{Author Contributions}
A.T. developed the algorithm, performed the numerical calculations, analysis and writing of the manuscript. All contributed to the discussions, analysis and writing of the manuscript. P.A.D. and S.T. supervised the project.

\subsection*{Acknowledgments}
Research presented in this article was supported by the Laboratory Directed Research and Development (LDRD) program of Los Alamos National Laboratory (LANL) under project number 20200056DR. This work was conducted in part at the Center for Integrated Nanotechnologies, a U.S. Department of Energy, Office of Basic Energy Sciences user facility.

\subsection*{Data Availability}
The data used to support the findings of this study are included within the article.

\section*{Supplementary Materials}

Supplementary materials contain additional results to support the findings of the main text. There are three parts.

\noindent Part 1 contains the results obtained using the TDHF (CIS) and the QAE running on the D-Wave 2000Q:

Fig. S1. Convergence of CIS excitation energies for the H$_2$ molecule with respect to basis set.

Fig. S2. CIS excitation energies for the umbrella inversion of ammonia (NH$_3$).

Table S1. Singlet CIS excitation energies (eV).

Table S2. Triplet CIS excitation energies (eV).

Table S3. Singlet CIS transition dipole moments (au).

Table S4. Singlet CIS oscillator strengths (unitless).

\noindent Part 2 repeats all TDDFT and CIS calculations, but with QUBOs solved classically:

Fig. S3. Classical convergence of TDDFT excitation energies for the H$_2$ molecule with respect to basis set.

Fig. S4. Classical convergence of CIS excitation energies for the H$_2$ molecule with respect to basis set.

Fig. S5. Classical TDDFT excitation energies for the umbrella inversion of ammonia (NH$_3$).

Fig. S6. Classical CIS excitation energies for the umbrella inversion of ammonia (NH$_3$). 

Table S5. Classical singlet TDDFT excitation energies (eV).

Table S6. Classical triplet TDDFT excitation energies (eV).

Table S7. Classical singlet TDDFT transition dipole moments (au).

Table S8. Classical singlet TDDFT oscillator strengths (unitless).

Table S9. Classical singlet CIS excitation energies (eV).

Table S10. Classical triplet CIS excitation energies (eV).

Table S11. Classical singlet CIS transition dipole moments (au).

Table S12. Classical singlet CIS oscillator strengths (unitless).

\noindent Part 3 demonstrates the heuristic nature of the qbsolv on the example of C$_2$H$_6$ in the STO-3G basis set.

Table S13. Classical singlet CIS/STO-3G excitation energies (eV) for C$_2$H$_6$, multiple attempts.

Table S14. Classical triplet CIS/STO-3G excitation energies (eV) for C$_2$H$_6$, multiple attempts.

\printbibliography

\end{document}